# The Mechanism behind the Embeddings of String Theories

José M. Figueroa-O'Farrill †

*Department of Physics, Queen Mary and Westfield College*
*Mile End Road, London E1 4NS, UK*

## ABSTRACT

It has been realised recently that there is no unique way to describe the physical states of a given string theory. In particular, it has been shown that any bosonic string theory can be embedded in a particular $N$=1 string background in such a way that the spectrum and the amplitudes of both theories agree. Similarly, it is also known that the amplitudes of any $N$=1 string theory can be obtained from a particular $N$=2 string background. When rephrased in the language of BRST cohomology, these results suggest a close connection to the theory of induced representations. The purpose of this note is to investigate this connection further and at the same time to reveal the mechanism behind these embeddings between string theories. We will first analyze the embedding of an affine algebra $\widehat{\mathfrak{g}}$ in the $N$=1 affine algebra associated to $\mathfrak{g}$. Given any BRST cohomology theory for $\widehat{\mathfrak{g}}$ we will be able to construct one for the $N$=1 affine algebra associated to $\mathfrak{g}$ such that the cohomologies agree as operator product algebras. This is proven in two different ways. This example is the simplest in its kind and, in a sense that is made precise in the paper, all other similar embeddings are deformations of this one. We conclude the paper with a brief treatment of the general case, where we prove that for a particular class of "good" embeddings, the cohomologies are again isomorphic.

† e-mail: jmf@strings1.ph.qmw.ac.uk

## §1 Introduction

One of the most recent surprises that string theory has had in store for us is that there is no unique way to describe the physical states of a given string theory. Take the "humble" bosonic string theory, for instance, with an arbitrary background. It was shown in [**1**] that one can cook up a background for the $N$=1 string whose physical states coincide with those of the original bosonic string background [**2**]. Similarly, it was further shown in [**1**] that any $N$=1 string background can be embedded in a particular $N$=2 string background in such a way that the amplitudes for the $N$=1 string can be recovered from $N$=2 string amplitudes, at least when the physical states come essentially from the matter sector. In this case the physical states have not been shown to coincide and, in fact, for that particular embedding it is not clear a priori that they should.

Let us rephrase the first of these embeddings in the language of BRST cohomology. Suppose that $\mathcal{B}$ is the Hilbert space of any bosonic string background; that is, of any conformal field theory with $c$=26. The physical space of the string theory is then the BRST cohomology $H_{N=0}(\mathcal{B})$. It was shown in [**1**] that adding a fermionic BC system of weights $(\frac{3}{2}, -\frac{1}{2})$, one can make a superconformal field theory with $c$=15. The Hilbert space of this superconformal field theory is the tensor product $\mathcal{B} \otimes \mathcal{F}$, where $\mathcal{F}$ is the Hilbert space of the conformal field theory corresponding to the BC system. Such a superconformal field theory can be interpreted as an $N$=1 string background, whose physical states are then given by the BRST cohomology $H_{N=1}(\mathcal{B} \otimes \mathcal{F})$. This $N$=1 string background is a very special background, however, in that all its supersymmetry is fake. In fact we showed in [**2**] that

$$H_{N=1}(\mathcal{B} \otimes \mathcal{F}) \cong H_{N=0}(\mathcal{B}) \ , \tag{1.1}$$

whence its physical states coincide with those of the original bosonic string theory, which has no supersymmetry at all. The proof in [**2**] is valid for arbitrary backgrounds $\mathcal{B}$, but has the drawback that the isomorphism was at the level of vector spaces; whence one could not conclude that the amplitudes agree. Soon after [**2**] was written, I found a modified proof [**3**] of (1.1) in such a way that the isomorphism now respects the operator product algebra (see below). More recently, however, a paper has appeared [**4**] in which an elementary and simple proof of (1.1) is given by conjugating the BRST differential of the $N$=1 theory to one which corresponds to a tensor product of the $N$=0 string with a topological theory. In other words, they lift the proof of (1.1) in [**2**] from the level of cohomology to (almost) the level of complexes and in this fashion make the isomorphism of operator product algebras manifest. It is not inconceivable that for the BRST cohomology theories that appear in physics, whenever two differentials compute the same cohomology, one can find a transformation relating one to the other (up to topological terms); but without general results guaranteeing its existence, one is left with the task of finding the explicit transformation. Whereas this is a relatively straightforward task (with or without computer) for the simplest examples, it is in general impracticable. On the other hand, the homological methods used in this paper are generally applicable; although—admittedly—not so elementary. At any rate, for the cohomology theories we deal with in this paper we will be able to use both approaches. More recent results in this field include the paper **[5]** in which the usual bosonic string spectrum can be recoverd from a particular $\mathsf{W}_3$ string; although it seems that the philosophy in that paper is somewhat different from the one spoused in the present paper.

As pointed out already in **[2]**, the above result (1.1) is very reminiscent of a similar result in the theory of induced representations, for which we refer the reader to **[6]**. Suppose that $\mathfrak{h} \subset \mathfrak{g}$ are Lie algebras and let $U(\mathfrak{h})$ and $U(\mathfrak{g})$ denote their universal enveloping algebras. Given a representation $V$ of $\mathfrak{h}$, there is a standard way to induce a representation of $\mathfrak{g}$. $V$ is in a natural way a left module over $U(\mathfrak{h})$, which basically means that we know how to multiply elements of $V$ on the left by elements of $U(\mathfrak{h})$. Since $\mathfrak{h}$ is a Lie subalgebra of $\mathfrak{g}$, then also $U(\mathfrak{h})$ is a subalgebra of $U(\mathfrak{g})$ and so $U(\mathfrak{g})$ is in a natural way a $U(\mathfrak{h})$-module. We can thus take the tensor product

$$W = U(\mathfrak{g}) \otimes_{U(\mathfrak{h})} V \ . \tag{1.2}$$

It is clear that $W$ is a module over $U(\mathfrak{h})$ but, more importantly, the left multiplication of $U(\mathfrak{g})$ on itself makes $W$ a left module over $U(\mathfrak{g})$, and hence it becomes a representation of $\mathfrak{g}$. The above construction is known as the induced module construction and $W$ is said to be induced from $V$ and one sometimes uses the notation $W = I_{\mathfrak{h}}^{\mathfrak{g}}(V)$ to make this manifest. Then analogue of (1.1) in this context is that in homology

$$H_*(\mathfrak{h}; V) \cong H_*(\mathfrak{g}; W) \ ; \tag{1.3}$$

or in other words, that all the homological information of $W$ is contained already in $V$.

Although (1.3) may not be so familiar to physicists, the induced module construction certainly is. Mathematically, it is but an instance of the "extension of scalars" by which, for example, a real vector space (that is, an $\mathbb{R}$-module) can be tensored with the complex numbers and hence be complexified. In physics, we are familiar with induced representations in at least two cases. In relativistic field theory, the unitary representations of the Poincaré group are induced from representations of the Lorentz group which are themselves induced from representations of the relevant "little group": $SU(2)$ for massive representations or the Euclidean group $E(2)$ for massless ones. This follows from the classical work of Wigner and Bargmann and later Mackey. Also, Verma modules of (Lie, W, ...) algebras are induced representations. For example, let $\mathsf{Vir}$ denote the Virasoro algebra and let us decompose it as $\mathsf{Vir} = \mathsf{Vir}_+ \oplus \mathsf{Vir}_0 \oplus \mathsf{Vir}_- = \mathsf{Vir}_{\geq 0} \oplus \mathsf{Vir}_-$ in the obvious manner. The Verma module $V(h,c)$ is defined as the free $\mathsf{Vir}_-$-module generated by the vector $|h,c\rangle$. This vector spans a one-dimensional representation of $\mathsf{Vir}_{\geq 0}$ which induces $V(h,c)$ à la (1.2).

The similarity between (1.1) and (1.3) is supplemented by the similarities to be found also in the construction of the induced module itself. The Poincaré–Birkhoff–Witt theorem tells us that any basis of $\mathfrak{g}$ gives rise to a basis of $U(\mathfrak{g})$. In fact, as vector spaces—and as $\mathfrak{g}$-modules—$U(\mathfrak{g})$ is isomorphic to the symmetric algebra $S(\mathfrak{g})$. In particular, the isomorphism $\mathfrak{g} \cong \mathfrak{h} \oplus (\mathfrak{g}/\mathfrak{h})$ gives rise to an isomorphism (as vector spaces, and as $\mathfrak{h}$-modules) $U(\mathfrak{g}) \cong S(\mathfrak{g}/\mathfrak{h}) \otimes U(\mathfrak{h})$. In other words, $U(\mathfrak{g})$ is a free $U(\mathfrak{h})$-module with basis $S(\mathfrak{g}/\mathfrak{h})$. Therefore as an $\mathfrak{h}$-module,

$$W = I_{\mathfrak{h}}^{\mathfrak{g}}(V) \cong S(\mathfrak{g}/\mathfrak{h}) \otimes V \ . \tag{1.4}$$

Of course, when we deal more generally with Lie superalgebras, then we have to modify these formulas slightly. The PBW theorem still applies but now we have to substitute the symmetric algebra $S(\mathfrak{g}/\mathfrak{h})$ by its supersymmetric counterpart; that is, we take the symmetric algebra of the even part of $\mathfrak{g}/\mathfrak{h}$ tensored with the exterior algebra of the odd part of $\mathfrak{g}/\mathfrak{h}$. Unfortunately I am not aware of a supersymmetric version of (1.3), but I would be surprised if it were not true.

In the case of the embedding of the bosonic string in the $N{=}1$ string, the relevant algebras are $\mathfrak{h} = \mathsf{Vir}$ and $\mathfrak{g} = \mathsf{Vir}_{N=1}$. Then $\mathfrak{g}/\mathfrak{h}$ would correspond to the modes of the supercurrent and $S(\mathfrak{g}/\mathfrak{h})$ would be the exterior algebra on these modes. This has the same number of degrees of freedom as the fermionic BC system that we must tensor by in order to get from the $N{=}0$ background to the $N{=}1$ background and the same statistics, but the similarity ends there. (1.3) holds in homology, but we are interested in BRST cohomology which has elements both of homology and of cohomology (see **[7]**). We can get a cohomological version of (1.3) by taking the dual modules $V^*$ and $W^*$. This brings the dual of $S(\mathfrak{g}/\mathfrak{h})$ into play. In our stringy example, this means that we expect the dual of the supercurrent to appear. As far as the Virasoro algebra is concerned—that is, as $\mathfrak{h}$-modules—the supercurrent is simply a primary field of weight $\frac{3}{2}$. Its dual representation would correspond to a primary field of weight $-\frac{1}{2}$. Intuitively then, one would expect a fermionic BC system of weights $(\frac{3}{2}, -\frac{1}{2})$ to be the CFT one tensors by, as was indeed found in **[1]**.

These remarks are meant to be primarily for motivation: no amount of heuristics can ever do justice to the precise treatment of the BRST analogue of (1.2) which, for the case of Lie algebras, can be found in Voronov's paper **[7]**. It follows from **[7]** that BRST cohomology is a derived functor and this fact yields a concise proof of (1.3) for BRST cohomology **[8]** much in the same style as the modern proof of (1.3) for the classical theory. From the point of view of string theory there is one major drawback in the homological approach of **[7]**: BRST cohomology is an operator product algebra, and we would want to extend the isomorphism (1.3) to that category. In other words, we are missing a CFT description of the induced module. We are also missing an extension to Lie superalgebras and to W-algebras. The extension to Lie superalgebras is not supposed to represent any difficulty **[8]** but the story is different for W-algebras, for which there is no general theory of BRST cohomology but rather a collection of examples.

The purpose of this note is to explore the CFT aspects of the induced module construction and in the process, we hope to reveal the mechanism behind the embeddings of string theories. We will analyze one example in great detail: the one associated to the embedding of an affine Lie algebra $\widehat{\mathfrak{g}}$ into the $N$=1 affine Lie algebra $S\widehat{\mathfrak{g}}$. The relevant concepts and notation are described in section 2, which also contains a CFT description of the induced module (1.2) in this case. In section 3 we prove the analogue of (1.3). The proof benefits from the existence of descent equations governing the BRST complex of $S\widehat{\mathfrak{g}}$ with values in the induced representation. Section 4 contains the proof that the isomorphism of the BRST cohomologies is one of operator product algebras. Section 5 discusses the picture-changing phenomenon and proves the isomorphism of the cohomology at each different picture. In the light of the results of **[4]** we give in section 6 a more elementary proof of the above results. We present both proofs because the techniques in sections 3 and 4 are more widely applicable, as evinced by the contents of section 7, where in conclusion we discuss more general embeddings.

## §2 The $\widehat{\mathfrak{g}}$ BRST cohomology

Let us now consider the simplest example of induced module construction in CFT. Let $\mathfrak{g}$ be a semisimple Lie algebra and let $\widehat{\mathfrak{g}}$ denote its affinization. We fix a basis $\{X_i\}$ for $\mathfrak{g}$ and a nondegenerate invariant bilinear form $\langle,\rangle$ with matrix $g_{ij} = \langle X_i\,, X_j\rangle$ relative to this basis. We also let the structure constants be given by $[X_i\,, X_j] = f_{ij}{}^k X_k$. We think of $\widehat{\mathfrak{g}}$ as represented by currents $J_i(z)$ satisfying the following OPEs:

$$J_i(z)J_j(w) = \frac{k g_{ij}}{(z-w)^2} + \frac{f_{ij}{}^k J_k(w)}{z-w} + \text{reg.} \tag{2.1}$$

In order to define the BRST operator associated to $\widehat{\mathfrak{g}}$ we need to introduce ghosts $(b_i, c^i)$ obeying

$$b_i(z)c^j(w) = \frac{\delta_i^j}{z-w} + \text{reg.} \tag{2.2}$$

The BRST operator $Q$ is the charge associated to the BRST current

$$J_{N=0} = J_i c^i - \tfrac{1}{2} f_{ij}{}^k b_k c^i c^j \ , \tag{2.3}$$

that is, $Q = \oint J_{N=0}(z)$. It is well-known that the BRST operator squares to zero if and only if $k = -c_{\mathfrak{g}}$, where $c_{\mathfrak{g}}$ is defined by $f_{ik}{}^l f_{jl}{}^k = c_{\mathfrak{g}} g_{ij}$. We will call a $\widehat{\mathfrak{g}}$-module $\mathfrak{M}$ *admissible* if it has $k = -c_{\mathfrak{g}}$. More generally, we will call a module over some algebra *admissible*, if it is such that the corresponding BRST operator squares to zero. For Lie algebras this simply means a cancellation of $c$-number anomalies, like the central charge for the case of (super)conformal algebras or the level for affine algebras. In other words, an admissible module is one for which the BRST cohomology is well defined. Strictly speaking, we should also demand that acting on the representation with $d$, we never encounter infinite sums of elements. Technically, this last condition is guaranteed by restricting $\mathfrak{M}$ to the category $\mathcal{O}$. Since in practice, all representations arising in conformal field theory are in this category, we will tacitly assume this condition in the definition of an admissible module. In the mathematical literature the BRST cohomology of an admissible module $\mathfrak{M}$ goes under the name of the semi-infinite cohomology of $\widehat{\mathfrak{g}}$ relative its center with coefficients in $\mathfrak{M}$.

### The induced representation and its BRST operator

Given an admissible $\widehat{\mathfrak{g}}$-module $\mathfrak{M}$, we can induce an admissible module $\mathfrak{N}$ over the $N$=1 affine Lie algebra $S\widehat{\mathfrak{g}}$ associated to $\mathfrak{g}$ in such a way that the $\widehat{\mathfrak{g}}$-cohomology of $\mathfrak{M}$ and the $S\widehat{\mathfrak{g}}$-cohomology of $\mathfrak{N}$ agree. The induced module is constructed by tensoring with the Hilbert space of fermionic BC systems $(\widetilde{b}_i, \widetilde{c}^i)$. These fields obey the OPEs (2.2). Consider then the following currents

$$I_i = J_i + f_{ij}{}^k \widetilde{b}_k \widetilde{c}^j \qquad \text{and} \qquad \psi_i = \widetilde{b}_i \ . \tag{2.4}$$

One readily verifies that $(I_i, \psi_i)$ obey

$$\begin{aligned} I_i(z)I_j(w) &= \frac{f_{ij}{}^k I_k(w)}{z-w} + \text{reg.} \\ I_i(z)\psi_j(w) &= \frac{f_{ij}{}^k \psi_k(w)}{z-w} + \text{reg.} \\ \psi_i(z)\psi_j(w) &= \text{reg.} \end{aligned} \tag{2.5}$$

which is nothing but $S\widehat{\mathfrak{g}}$ at level zero. Happily, this is precisely the level at which the BRST cohomology of $S\widehat{\mathfrak{g}}$ makes sense.

Let us now write down the BRST operator for $S\widehat{\mathfrak{g}}$. To this end we introduce fermionic ghosts $(b_i, c^i)$ and bosonic ones $(\beta_i, \gamma^i)$ associated to the generators $I_i$ and $\psi_i$ respectively. The BRST operator is then the charge of the BRST current:

$$J_{N=1} = I_i c^i + \psi_i \gamma^i + f_{ij}{}^k \beta_k \gamma^j c^i - \tfrac{1}{2} f_{ij}{}^k b_k c^i c^j \ , \tag{2.6}$$

which in the particular representation defined by (2.4) becomes

$$J_{N=1} = \widetilde{b}_i \gamma^i + J_i c^i + f_{ij}{}^k \beta_k \gamma^j c^i + f_{ij}{}^k \widetilde{b}_k \widetilde{c}^j c^i - \tfrac{1}{2} f_{ij}{}^k b_k c^i c^j \ . \tag{2.7}$$

We shall let $D$ denote the corresponding BRST charge.

The complex

Let us denote by $C$ the complex in which $D$ acts. It is given by

$$C = \mathfrak{N} \otimes \mathcal{H}_{bc} \otimes \mathcal{H}_{\beta\gamma} \tag{2.8}$$

where $\mathcal{H}_{bc}$ and $\mathcal{H}_{\beta\gamma}$ are the Hilbert spaces of the $(b_i, c^i)$ and $(\beta_i, \gamma^i)$ ghost systems, respectively; and $\mathfrak{N}$ is the $S\widehat{\mathfrak{g}}$-module induced from $\mathfrak{M}$. In other words, as a $\widehat{\mathfrak{g}}$-module, $\mathfrak{N} = \mathfrak{M} \otimes \mathcal{H}_{\widetilde{b}\widetilde{c}}$, with $\mathcal{H}_{\widetilde{b}\widetilde{c}}$ the Hilbert space of the $(\widetilde{b}_i, \widetilde{c}^i)$ BC system. There are two caveats we must worry about. First of all, in choosing the complex we have to distinguish between the NS and R sectors for the fields with half-integer weight. Moreover because we are dealing with bosonic BC systems, we must distinguish between all the different inequivalent representations; that is, between the different pictures. For the ease of exposition we will restrict ourselves to the NS sector and, at first, to the picture corresponding to the projective-invariant vacuum. We will see in section 5 that the cohomology for any other picture is isomorphic to this one—the isomorphism being given by a picture-changing operator.

In the NS sector the fields are single-valued and thus fields with half-integral weights $(\widetilde{b}_i, \widetilde{c}^i, \beta_i, \gamma^i)$ are half-integrally moded. We now define the Hilbert space of our BC systems in more detail. Let us define the vacuum $|0\rangle_{bc}$ by

$$\begin{aligned} b_{i,n}|0\rangle_{bc} &= 0 \qquad \forall i \quad \forall n \geq 1 \\ \text{and } c^i_n|0\rangle_{bc} &= 0 \qquad \forall i \quad \forall n \geq 0 \ . \end{aligned} \tag{2.9}$$

Similarly, we define $|0\rangle_{\beta\gamma}$ by

$$\begin{aligned} \beta_{i,n}|0\rangle_{\beta\gamma} &= 0 \qquad \forall i \quad \forall n \geq \tfrac{1}{2} \\ \text{and } \gamma^i_n|0\rangle_{\beta\gamma} &= 0 \qquad \forall i \quad \forall n \geq \tfrac{1}{2} \ ; \end{aligned} \tag{2.10}$$

and

$$\begin{aligned} \widetilde{b}_{i,n}|0\rangle_{\widetilde{b}\widetilde{c}} &= 0 \qquad \forall i \quad \forall n \geq \tfrac{1}{2} \\ \text{and } \widetilde{c}^i_n|0\rangle_{\widetilde{b}\widetilde{c}} &= 0 \qquad \forall i \quad \forall n \geq \tfrac{1}{2} \ . \end{aligned} \tag{2.11}$$

The Hilbert spaces are then obtained by taking linear combinations of monomials obtained by acting on the relevant vacuum with the remaining modes. The Hilbert spaces of the ghost systems come graded by the ghost number, which is the eigenvalue of the charge associated to the chiral currents $-b_i c^i$ and $-\beta_i \gamma^i$. The vacua have zero ghost number. The operator product preserves both ghost numbers separately, and we will it find it instrumental to keep track of them.

## §3 The Descent Equations

In this section we will compute the cohomology of the operator $D$ using some primitive homological techniques. We will show that $D$ can be understood as the total differential of a double complex. We will exhibit some descent equations governing this double complex and use this fruitfully to compute the cohomology of $D$.

The double complex

Let us define $C^{p,q}$ to be the subspace of $C$ with $(b,c)$-ghost number $p$ and $(\beta,\gamma)$-ghost number $q$. That is,

$$C^{p,q} = \mathfrak{N} \otimes \mathcal{H}^p_{bc} \otimes \mathcal{H}^q_{\beta\gamma} \ . \tag{3.1}$$

We will say that elements of $C^{p,q}$ have bidegree $(p,q)$. Relative to this bidegree the BRST operator D breaks up into two pieces $D = d + \delta$, of bidegrees $(0,1)$ and $(1,0)$ respectively. Explicitly,

$$d = \oint \widetilde{b}_i \gamma^i \tag{3.2}$$

and

$$\delta = \oint \left( J_i c^i + f_{ij}{}^k \beta_k \gamma^j c^i + f_{ij}{}^k \widetilde{b}_k \widetilde{c}^j c^i - \tfrac{1}{2} f_{ij}{}^k b_k c^i c^j \right) \ . \tag{3.3}$$

Because the algebra preserves the bidegree, we can break up the equations $D^2 = 0$ into terms of different bidegree to obtain $d^2 = \delta^2 = [d\,, \delta] = 0$. In other words we have a double complex, which can be represented pictorially as a grid, with $d$ and $\delta$ acting vertically and horizontally respectively:

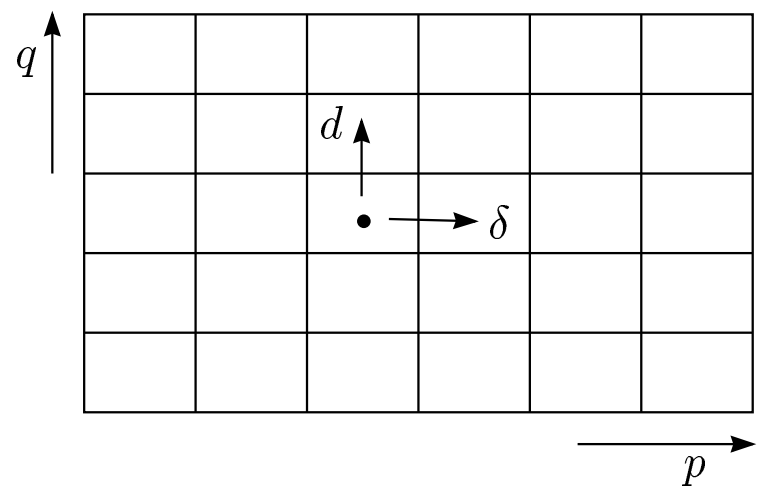

There are standard ways to attack the computation of the cohomology of a double complex which involves a rather heavy machinery known as a spectral sequence. But in this case we can do things by hand.

The cohomology of $d$

The first thing to notice is that the cohomology of $d$ is very simple to compute. First of all, $d : C^{p,q} \to C^{p,q+1}$ and in fact only acts on $\mathcal{H}_{\widetilde{b}\widetilde{c}} \otimes \mathcal{H}_{\beta\gamma}$. It was shown in **[9]** (see the proof of the lemma in Section 3) that there exists an operator $K$ such that $[d\,, K] = N$, where $N$ is the total occupation number operator in $\mathcal{H}_{\widetilde{b}\widetilde{c}} \otimes \mathcal{H}_{\beta\gamma}$. Explicitly, the operator $K$ is given by

$$K = \sum_{r\geq 1/2} \left(\beta_{i,-r}\widetilde{c}^i_r - \widetilde{c}^i_{-r}\beta_{i,r}\right) \ , \tag{3.4}$$

and the total occupation number operator is given by $N = N_{\beta\gamma} + N_{\widetilde{b}\widetilde{c}}$ with

$$N_{\beta\gamma} = \sum_{r\geq 1/2} \left(\beta_{i,-r}\gamma^i_r - \gamma^i_{-r}\beta_{i,r}\right) \tag{3.5}$$

and

$$N_{\widetilde{b}\widetilde{c}} = \sum_{r\geq 1/2} \left(\widetilde{b}_{i,-r}\widetilde{c}^i_r + \widetilde{c}^i_{-r}\widetilde{b}_{i,r}\right) \tag{3.6}$$

Thus $N$ is chain homotopic to zero, whence all cohomology occurs in the subspace annihilated by $N$. But this subspace is one-dimensional: in fact, it is spanned by the vacuum $|0\rangle_{\widetilde{b}\widetilde{c}} \otimes |0\rangle_{\beta\gamma}$; hence the cohomology of $d$ is given by $\mathfrak{M} \otimes \mathcal{H}_{bc} \otimes |0\rangle_{\widetilde{b}\widetilde{c}} \otimes |0\rangle_{\beta\gamma} \cong \mathfrak{M} \otimes \mathcal{H}_{bc}$ which agrees as a vector space with the $\widehat{\mathfrak{g}}$ BRST complex for the representation $\mathfrak{M}$. In particular, all the $d$-cohomology is concentrated in the row $q = 0$ of our double complex.

The descent equations

We now investigate the horizontal differential $\delta$ given by (3.3). Notice that it can be written as $\delta = Q + [d\,, X]$ where $Q$ is the $\widehat{\mathfrak{g}}$ BRST differential given by the integral of (2.3) and $X$ is given by $X = \oint f_{ij}{}^k \beta_k \widetilde{c}^j c^i$. We claim that the cohomology of $Q$ and the cohomology of $D$ agree. The proof consists in showing that every $Q$-cocycle gives rise to a $D$-cocycle and vice versa and that this map sends coboundaries to coboundaries. This will then allow us to conclude that the cohomologies are isomorphic. We will present the correspondence between cocycles in some detail, leaving the one between coboundaries as an exercise to the diligent reader.

Suppose that $\psi \in \mathfrak{M} \otimes \mathcal{H}^p_{bc}$ is a $Q$-cocycle. Let us consider $\psi_0 \in C^{p,0}$ obtained from $\psi$ by tensoring with the vacuum $|0\rangle_{\widetilde{b}\widetilde{c}} \otimes |0\rangle_{\beta\gamma}$; that is, $\psi_0 = \psi \otimes |0\rangle_{\widetilde{b}\widetilde{c}} \otimes |0\rangle_{\beta\gamma}$. Then it is clear that $d\psi_0 = 0$ and also that $\delta\psi_0 = dX\psi_0$.

Put $\psi_{-1} = -X\psi_0$ and consider $\psi_0 + \psi_{-1}$. It is in general not a $D$-cocycle, but instead $D(\psi_0 + \psi_{-1}) = \delta\psi_{-1} \in C^{p+2,-1}$. Now, $d\delta\psi_{-1} = -\delta d\psi_{-1} = \delta dX\psi_0 = \delta^2\psi_0 = 0$; whence $\delta\psi_{-1}$ is a $d$-cocycle. However since it lies outside the region where the $d$-cohomology lives, we know that it is a coboundary: $\delta\psi_{-1} = -d\psi_{-2}$, for some $\psi_{-2} \in C^{p+2,-2}$. Continuing in this fashion, we arrive at a $D$-cocycle $\Psi = \psi_0 + \psi_{-1} + \psi_{-2} + \cdots$ with $\psi_{-n} \in C^{p+n,-n}$:

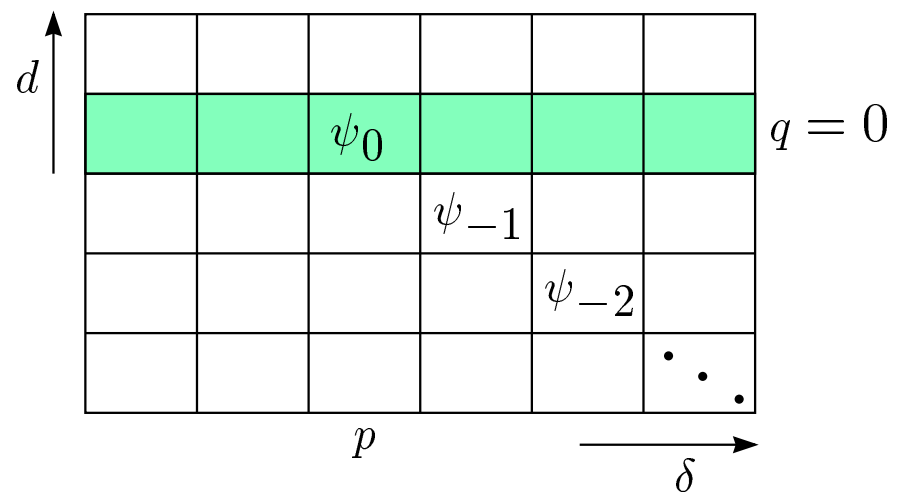

This series actually terminates because the complex is naturally graded by the eigenvalues of the analog of the Virasoro generator $L_0$ which counts the moding of our basic fields. In other words, we define $L_0$ as follows. If $\phi(z)$ is any field, its mode $\phi_n$ has weight $-n$: $[L_0\,, \phi_n] = -n\phi_n$. Because $D$ commutes with $L_0$, we can work in one eigenspace of $L_0$ at a time and simply collate the results at the end of the day. It is then straightforward to verify that for a fixed eigenvalue of $L_0$ and for a fixed total ghost number $p + q$, the number of nonzero $C^{p,q}$ is actually finite.

Conversely, if $\Psi$ is any $D$-cocycle, we can show it to be $D$-cohomologous to a cocycle of the form $\psi_0 + \psi_1 + \cdots$ with $\psi_0 = \psi \otimes |0\rangle_{\widetilde{b}\widetilde{c}} \otimes |0\rangle_{\beta\gamma}$ and $Q\psi = 0$. Indeed, suppose that $D\Psi = 0$ and that $\Psi$ has total ghost number $p$. Then we can break it up as $\Psi = \psi_N + \psi_{N-1} + \cdots + \psi_0 + \psi_{-1} + \cdots$ with again $\psi_n \in C^{p-n,n}$. Assume first that $N > 0$. Notice that since $D\Psi = 0$, $\psi_N = 0$ is a $d$-cocycle. Since there is no $d$-cohomology for $q = N > 0$ it is also a $d$-coboundary: $\psi_N = d\xi_{N-1}$. Consider then $\Psi - D\xi_{N-1}$. It is a $D$-cocycle cohomologous to $\Psi$ but such that it has the form $\psi'_{N-1} + \cdots$. Repeating the procedure it is clear that there exists $\Xi = \xi_{N-1} + \cdots + \xi_0$ such that $\Psi - D\Xi = \psi''_0 + \cdots$:

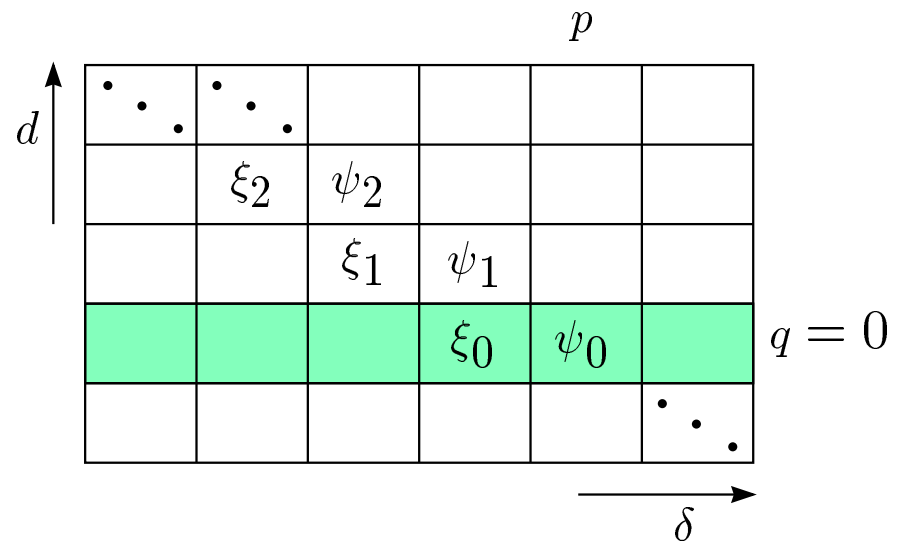

Again it follows that $\psi_0''$ is a $d$-cocycle. Since it has $q=0$ it is not necessarily a $d$-coboundary, but we can decompose it as follows: $\psi_0''=\psi\otimes|0\rangle_{\widetilde{b}\widetilde{c}}\otimes|0\rangle_{\beta\gamma}+d\xi_{-1}$, for some $\xi_{-1}$. In other words, $\widetilde{\Psi}\equiv\Psi-D(\xi_{N-1}+\cdots+\xi_{-1})=\psi\otimes|0\rangle_{\widetilde{b}\widetilde{c}}\otimes|0\rangle_{\beta\gamma}+\cdots$, where the $\cdots$ indicate terms of $q<0$. We claim that $\psi$ is the desired $Q$-cocycle. Let $\widetilde{\Psi}=\widetilde{\psi}_0+\widetilde{\psi}_1+\cdots$. Since $D\widetilde{\Psi}=0$, it follows that $\delta\widetilde{\psi}_0=-d\widetilde{\psi}_1$; or, using that $\delta=Q+[d\,,X]$, that

$$Q\widetilde{\psi}_0+dX\widetilde{\psi}_0=-d\widetilde{\psi}_1\ . \tag{3.7}$$

In other words,

$$Q\psi\otimes|0\rangle_{\widetilde{b}\widetilde{c}}\otimes|0\rangle_{\beta\gamma}=-d(X\widetilde{\psi}_0+\widetilde{\psi}_1)\ . \tag{3.8}$$

But the left-hand-side of the equation is manifestly a nontrivial $d$-cocycle, whereas the right-hand-side is a $d$-coboundary. This can only be true if both are separately zero. Thus $Q\psi=0$ as advertised.

The upshot of all this is that we have exhibited a bijective correspondence between $Q$-cocycles and $D$-cocycles. It is moreover easy to verify that this correspondence relates coboundaries to coboundaries and it is therefore well-defined in cohomology. We can therefore conclude that the cohomology of $D$ and that of $Q$ are isomorphic; equivalently that the BRST cohomology of $\widehat{\mathfrak{g}}$ with coefficients in $\mathfrak{M}$ and that of $S\widehat{\mathfrak{g}}$ with coefficients in $\mathfrak{N}$ (in the NS sector and in the 0th picture) are isomorphic as vector spaces:

$$H(\widehat{\mathfrak{g}},\mathfrak{M})\cong H(S\widehat{\mathfrak{g}},\mathfrak{N})\ . \tag{3.9}$$

This is not the end of the story, however, since the isomorphism also preserves the algebraic structures defined by the operator product expansion.

## §4 Cohomology isomorphism as operator product algebras

In CFT there is a one-to-one correspondence between states and fields. Given a state $\psi$ there is a field $\psi(z)$ which creates it when acting on the projective-invariant vacuum at $z=0$. This correspondence extends to one between BRST-invariant states and BRST-invariant fields: $Q\psi=0$ if and only if $[Q\,,\psi(z)]=0$, using the fact that the BRST operator annihilates the vacuum. It is also clear that a similar correspondence exists between BRST trivial states and BRST trivial fields: $(Q\xi)(z)=[Q\,,\xi(z)]$. Therefore we can describe the BRST cohomology either with states or with fields. One advantage of the field-theoretic description is that it allows us to define algebraic operations in cohomology.

If $A(z)$ and $B(z)$ are two conformal fields, their OPE defines a family of bilinear operations $[AB]_n(z)$ by

$$A(z)B(w)=\sum_{n<<\infty}\frac{[AB]_n(w)}{(z-w)^n}\ . \tag{4.1}$$

These brackets obey some identities that have been recently formalized by mathematicians (see **[10]** for a brief review). I call such a structure generically an operator product algebra (OPA) to distinguish it from the vertex operator algebra (VOA), which assumes the existence of a Virasoro subalgebra. Other popular nomenclature includes vertex algebra, and pre-vertex operator algebra.

If we let $J(z)$ denote a generic BRST current and $Q$ its charge, then we have that $[Q\,,A(z)]=[JA]_1(z)$, from where it follows, using the associativity of the OPE, that the operation $A(z)\mapsto[JA]_1(z)$ is a derivation over all of the brackets; in other words,

$$[J[AB]_n]_1(z)=[[JA]_1B]_n(z)+(-)^{|A|}[A[JB]_1]_n(z)\qquad\forall n\ . \tag{4.2}$$

It follows from this relation that if $A(z)$ and $B(z)$ are BRST-invariant fields, so are all their brackets $[AB]_n(z)$. Similarly, if $A(z)$ is BRST-invariant and $B(z)=[Q\,,C(z)]$ is BRST trivial, then so are all the brackets $[AB]_n(z)$. In other words, the brackets descend to cohomology making it into an operator product algebra.

If to both cohomology spaces in (3.9) we now give the structure of operator product algebras which follows from the above discussion, it is natural to ask if the isomorphism extends to an isomorphism of OPAs. Not surprisingly, perhaps, the answer is affirmative. To summarize the proof in a nutshell, the crucial fact is that $D$-cocycles are uniquely determined (via descent) by their $q=0$ piece which can be chosen to be a $Q$-cocycle, as we will see. Therefore the brackets on $Q$-cocycles determine uniquely the brackets on $D$-cocycles. We now fill the details.

Suppose that $\psi$ and $\phi$ are $Q$-cocycles at $(b,c)$-ghost numbers $p$ and $p'$ respectively. The associated fields obey $[Q\,,\psi(z)]=[Q\,,\phi(z)]=0$. Let $\Psi=\psi_0+\psi_{-1}+\cdots$ and $\Phi=\phi_0+\phi_{-1}+\cdots$ be the corresponding $D$-cocycles, with $\psi_{-n}\in C^{p+n,-n}$ and $\phi_{-n}\in C^{p'+n,-n}$; and let $\Psi(z)$ and $\Phi(z)$ be the associated $D$-invariant fields. Since the field associated to the vacuum is the identity, notice that $\psi_0(z)=\psi(z)$ and $\phi_0(z)=\phi(z)$. Let us denote by $\iota$ the map $\iota:\Psi(z)\mapsto\psi(z)$. We want to show that it is an operator algebra homomorphism; that is,

$$\iota([\Psi\Phi]_n(z))=[\iota(\Psi)\iota(\Phi)]_n(z)=[\psi\phi]_n(z)\ . \tag{4.3}$$

Because the operator product preserves the $(p,q)$ bidegree, we can decompose the brackets $[\Psi\Phi]_n(z)$ into terms of different bidegree. It is clear that $[\Psi\Phi]_n(z)$ has total ghost number $p+p'$ and it will contain pieces of bidegrees $(p+p'+i, -i)$ for $i = 0, 1, 2, \ldots$. In particular the piece of bidegree $(p+p', 0)$ comes from the bracket of the $(p, 0)$ piece of $\Psi(z)$ with the $(p', 0)$ piece of $\Phi(z)$:

| | | | | | | | |
|---|---|---|---|---|---|---|---|
| | | | | | | | |
| $\psi_0$ | | $\phi_0$ | | | $\cdots$ | $[\psi_0\phi_0]_n$ | |
| | $\psi_{-1}$ | | $\phi_{-1}$ | | | | $\ddots$ |
| | | $\psi_{-2}$ | | $\phi_{-2}$ | | | |
| | | | $\ddots$ | | | | |
| $p$ | | $p'$ | | | | $p+p'$ | |

But $[\psi_0\phi_0]_n(z) = [\psi\phi]_n(z)$. Therefore, $[\Psi\Phi]_n(z) = [\psi\phi]_n(z) + \cdots$ which means that $\iota([\Psi\Phi]_n(z)) = [\psi\phi]_n(z)$ which is precisely (4.3). Hence, $\iota$ which was previously shown to be a vector space isomorphism is now shown to also preserve the OPA structure.

## §5 Cohomology in other pictures: picture-changing operators

Whenever we have bosonic ghosts we need to consider all the inequivalent representations of the mode algebra. This is most transparent from the bosonized form of the bosonic BC system **[11]**. Introducing scalar bosons $\varphi_i(z)$ and fermionic BC systems $(\eta_i, \xi_i)$ obeying the OPEs

$$\varphi_i(z)\varphi_j(w) = -\delta_{ij}\log(z-w) + \cdots \tag{5.1}$$

and

$$\eta_i(z)\xi_j(w) = \frac{\delta_{ij}}{z-w} + \cdots \ , \tag{5.2}$$

we can then write

$$\gamma^i = e^{\varphi_i}\eta_i \qquad \text{and} \qquad \beta_i = e^{-\varphi_i}\partial\xi_i \ . \tag{5.3}$$

We can further bosonize the fermionic $(\eta_i, \xi_i)$ systems by introducing further bosons $\chi_i$ obeying $\chi_i(z)\chi_j(w) = \delta_{ij}\log(z-w) + \cdots$ in such a way that

$$\eta_i = e^{-\chi_i} \qquad \text{and} \qquad \xi_i = e^{\chi_i} \ . \tag{5.4}$$

Composing (5.3) with (5.4) yields an embedding of the operator algebra generated by $(\beta_i, \gamma^i)$ into the one generated by $(\varphi_i, \chi_i)$ and as a result an irreducible representation of the latter algebra will generally decompose into many irreducible representations of the latter. This is how the phenomenon of "pictures" arises. One notices that the charge associated to the combined $U(1)$ current $\partial\varphi_i + \partial\chi_i$ commutes with both $\beta_i$ and $\gamma^i$ since they depend on $\varphi_i - \chi_i$ and therefore indexes the irreducible representations of their algebra. Hereafter we will refer to the algebra generated by the $(\beta_i, \gamma^i)$ as the "small" algebra to distinguish it from the "large" algebra generated by $(\varphi_i, \chi_i)$ or equivalently by $(\varphi_i, \eta_i, \xi_i)$. Another way to understand the embedding of the small algebra into the large one, is to notice that the zero modes of the $\xi_i$ never appear in the expressions for $\beta_i$ or $\gamma^i$ since these depend on $\partial\xi_i$ and not on $\xi_i$ itself.

The most significant property of the large algebra is that it trivializes the cohomology. To see this consider the differential $d$ of (3.2). It is a sum of commuting differentials and the Künneth theorem tells us that the cohomology will be given by the tensor product of the cohomology of each individual differential. We will therefore consider one such differential $\widetilde{d} = \oint \widetilde{b}\gamma$. Using (5.3) it can be written in partially bosonized form as

$$\widetilde{d} = \oint \widetilde{b}e^{\varphi}\eta \ . \tag{5.5}$$

We claim that it is chain homotopic to the identity in the large algebra; in other words, that there is an operator $K$ in the large algebra such that $[\widetilde{d}, K] = 1$. Indeed, let $k(z) = (\widetilde{c}e^{-\varphi}\xi)(z)$. It is then easy to prove that

$$(\widetilde{b}e^{\varphi}\eta)(z)(\widetilde{c}e^{-\varphi}\xi)(w) = \frac{1}{z-w} + \text{reg.} \ ; \tag{5.6}$$

whence if $K = \oint k(z)$, $[\widetilde{d}, K] = 1$ as advertised. Hence the cohomology of $d$ is trivial, being the tensor product of trivial cohomologies.

This implies that the cohomology of $D$ is also trivial. For suppose $\Psi = \psi_N + \psi_{N-1} + \cdots$ is a $D$-cocycle. Then in particular $\psi_N$ is a $d$-cocycle. Since the $d$ cohomology is trivial, $\psi_N = d\xi_{N-1}$ for some $\xi_{N-1}$. Therefore $\Psi - D\xi_{N-1} = \psi'_{N-1} + \cdots$. This is still a $D$-cocycle and we can repeat the same procedure. Eventually, since the complex is levelwise finite, the process will terminate and $\Psi = D\Xi$ for some $\Xi = \xi_{N-1} + \cdots$. Hence every $D$-cocycle is a $D$-coboundary.

By an argument of Narganes-Quijano **[12]** this can be shown to lead at once to the existence of picture changing operators that relate the cohomology of $D$ in the small algebra from one picture to the next. We again present the argument for the differential $\widetilde{d}$. Let $\mathcal{A}$ stand for the small algebra of fields $(\widetilde{b}, \widetilde{c}, \varphi, \eta, \partial\xi)$, and let $\overline{\mathcal{A}}$ denote the larger algebra generated by $(\widetilde{b}, \widetilde{c}, \varphi, \eta, \xi)$. Any field in the small algebra can be thought of a particular kind of field of the large algebra, whence we have a natural embedding $i : \mathcal{A} \to \overline{\mathcal{A}}$. On the other hand, every field in the large algebra can be written as $\psi(z) + (\xi\phi)(z)$ with $\psi(z)$ and $\phi(z)$ fields in the small algebra. So we have a projection $p : \overline{\mathcal{A}} \to \mathcal{A}$ given by sending $\psi(z) + (\xi\phi)(z)$ to $\phi(z)$. It is clear then that the following sequence is exact:

$$0 \longrightarrow \mathcal{A} \overset{i}{\longrightarrow} \overline{\mathcal{A}} \overset{p}{\longrightarrow} \mathcal{A} \longrightarrow 0 \ . \tag{5.7}$$

Furthermore, both $i$ and $p$ (anti)commute with $\widetilde{d}$. This is obvious for $i$, whereas for $p$ we simply notice that

$$[\widetilde{d}, \xi\phi] = [\widetilde{d}, \xi]\phi - \xi[\widetilde{d}, \phi] \ , \tag{5.8}$$

and that since $[\widetilde{d}, \xi] = \widetilde{b}e^{\varphi}$, then the first term in the RHS of (5.8) is actually in the small algebra. Or said differently, $p$ anticommutes with $[\widetilde{d}, \cdot]$:

$$p([\widetilde{d}, \xi\phi]) = -[\widetilde{d}, \phi] = -[\widetilde{d}, p(\xi\phi)] \ .$$

This means that (5.7) is actually an exact sequence of complexes yielding a long exact sequence in cohomology:

$$\begin{array}{l} \cdots \longrightarrow H(\mathcal{A}) \longrightarrow H(\overline{\mathcal{A}}) \longrightarrow H(\mathcal{A}) \\ \qquad \xrightarrow{\widetilde{d}^*} H(\mathcal{A}) \longrightarrow H(\overline{\mathcal{A}}) \longrightarrow H(\mathcal{A}) \longrightarrow \cdots \end{array} \tag{5.9}$$

where every third term is zero, since $H(\overline{\mathcal{A}}) = 0$. In other words, the Bockstein map $\widetilde{d}^*$ defines an isomorphism $\widetilde{d}^* : H(\mathcal{A}) \to H(\mathcal{A})$. But what is the Bockstein map in this instance? It is actually defined as follows. Suppose $\phi(z)$ is a field in the small algebra obeying $[\widetilde{d}, \phi(z)] = 0$, then its preimage under $p$ is $(\xi\phi)(z)$. Now, $[\widetilde{d}, (\xi\phi)(z)] = ([\widetilde{d}, \xi]\phi)(z)$, which is a $\widetilde{d}$-invariant field in the small algebra. The cohomology class of this cocycle is the image of the Bockstein map. In other words, it is represented by normal ordering with the field $[\widetilde{d}, \xi(z)] = (\widetilde{b}e^{\varphi})(z)$. But notice that this map raises the $(\varphi + \chi)$-charge by one, hence it maps the cocycles in one picture to cocycles in the next. It is, in fact, the picture-changing operator $P$. Let us see how it acts. In the zeroth picture, the only nontrivial cocycle is the projective-invariant vacuum $|0\rangle_{\widetilde{b}\widetilde{c}} \otimes |0\rangle_{\beta\gamma}$. Its image under the picture-changing operator $P$ is

$$P(|0\rangle_{\widetilde{b}\widetilde{c}} \otimes |0\rangle_{\beta\gamma}) = \widetilde{b}(0)|0\rangle_{\widetilde{b}\widetilde{c}} \otimes (e^{\varphi})(0)|0\rangle_{\beta\gamma} = |-1\rangle_{\widetilde{b}\widetilde{c}} \otimes |1\rangle_{\beta\gamma} \ , \tag{5.10}$$

where the $q$-vacua are defined as follows:

$$\begin{aligned} \widetilde{b}_n |q\rangle_{\widetilde{b}\widetilde{c}} &= 0 \qquad \forall n \geq \tfrac{1}{2} + q \\ \widetilde{c}_n |q\rangle_{\widetilde{b}\widetilde{c}} &= 0 \qquad \forall n \geq \tfrac{1}{2} - q \ , \end{aligned} \tag{5.11}$$

and

$$\begin{aligned} \beta_n |q\rangle_{\beta\gamma} &= 0 \qquad \forall n \geq \tfrac{1}{2} - q \\ \gamma_n |q\rangle_{\beta\gamma} &= 0 \qquad \forall n \geq \tfrac{1}{2} + q \ . \end{aligned} \tag{5.12}$$

The different vacua for the fermionic $(\widetilde{b}, \widetilde{c})$-system are connected to each other by the action of a finite number of the modes; for example,

$$|-1\rangle_{\widetilde{b}\widetilde{c}} = \widetilde{b}_{-1/2}|0\rangle_{\widetilde{b}\widetilde{c}} \qquad \text{and} \qquad |1\rangle_{\widetilde{b}\widetilde{c}} = \widetilde{c}_{-1/2}|0\rangle_{\widetilde{b}\widetilde{c}} \ . \tag{5.13}$$

On the other hand, the vacua for the bosonic $(\beta, \gamma)$-system define inequivalent representations of the operator algebra. In fact, to connect them one must use the normal-ordered exponentials of $\varphi$:

$$|q\rangle_{\beta\gamma} = (e^{q\varphi})(0)|0\rangle_{\beta\gamma} \ . \tag{5.14}$$

Now, applying the picture-changing operator a second time, we find that

$$P^2(|0\rangle_{\widetilde{b}\widetilde{c}} \otimes |0\rangle_{\beta\gamma}) = |-2\rangle_{\widetilde{b}\widetilde{c}} \otimes |2\rangle_{\beta\gamma} \ , \tag{5.15}$$

where we have used that $(\widetilde{b}e^{\varphi})(\widetilde{b}e^{\varphi}) = \partial\widetilde{b}\widetilde{b}e^{2\varphi}$. Continuing in this fashion it is easy to prove that

$$P^q(|0\rangle_{\widetilde{b}\widetilde{c}} \otimes |0\rangle_{\beta\gamma}) = |-q\rangle_{\widetilde{b}\widetilde{c}} \otimes |q\rangle_{\beta\gamma} \ , \qquad \text{for } q \geq 0. \tag{5.16}$$

Notice that the $q$-vacuum has charge $q$, which for the $(\beta, \gamma)$-system gets translated to ghost number $q$. The picture-changing operator is invertible, since it defines an isomorphism in cohomology. Therefore, we can use its inverse $P^{-1}$ to create cohomology classes of negative ghost number. One can work out the expression for $P^{-1}$ explicitly and it turns out that $(P^{-1}\psi)(z) = (\widetilde{c}e^{-\varphi}\psi)(z)$. In particular, (5.16) is now true for $q < 0$. In summary, we have one nontrivial cocycle at each ghost number.

In our particular cohomology problem we have $\dim \mathfrak{g}$ picture-changing operators $P_i$. Thus the cohomology of $d$ is highly degenerate. Nevertheless if we fix one particular irreducible representation of the $(\beta_i, \gamma^i)$-algebra, then there is precisely one nontrivial cocycle. One can now take, in any picture, the same steps that led to the descent equations described above for the zeroth picture. This allows us to conclude that the $D$-cohomology in any one picture is isomorphic to the one in the zeroth picture; although the isomorphism will be given by different picture-changing operators $\widetilde{P}_i$.

## §6 A more economical proof

In the light of the work of **[4]** on the $N$=0 $\subset$ $N$=1 string embeddings, one may wonder whether the result that has just been described could not have been obtained in a more economical fashion by conjugating the BRST operator into one whose cohomology could be seen *manifestly* be the one we computed. In other words, can we cook up a weight zero charge $R = \oint r(z)$ in such a way that

$$e^{R}(d+Q)e^{-R} = D \ ? \tag{6.1}$$

Let us first see what (6.1) entails. Notice that

$$\begin{aligned} e^{R}(d+Q)e^{-R} &= \exp \operatorname{ad} R \cdot (d+Q) \\ &= d + [R,d] + Q + \tfrac{1}{2}[R,[R,d]] + [R,Q] + \cdots \ . \end{aligned} \tag{6.2}$$

The fact that $\delta = Q + [d\,,X]$ suggests that we try $R = -X = \oint f_{ij}{}^{k}\beta_k c^i \widetilde{c}^j$ at least as a first approximation. Notice that $X$ has bidegree $(1,-1)$, and that $Q$ and $[d,X]$ have the same bidegree $(1,0)$. Therefore the terms $(\operatorname{ad} X)^{n+1}\cdot d$ and $(\operatorname{ad} X)^{n}\cdot Q$ will have the same bidegree: $(n+1,-n)$. Let us analyze the equation

$$e^{-X}(d+Q)e^{X} - D = \tfrac{1}{2}[X,[X,d]] - [X,Q] - \tfrac{1}{6}[X,[X,[X,d]]] + \tfrac{1}{2}[X,[X,Q]] + \cdots \tag{6.3}$$

one bidegree at a time. The terms of bidegree $(2,-1)$ cancel as a result of the Jacobi identity; whence we find

$$[Q,X] = -\tfrac{1}{2}[[d,X],X] \ . \tag{6.4}$$

Plugging this back into (6.3) we find that the contribution to bidegree $(3,-2)$ is simply $-\frac{1}{12}[[[d,X],X],X]$. But this is automatically zero because there is nothing that can contract. Now, the other terms are also zero because as a result of (6.4) we can express all the other terms in terms of $(\operatorname{ad} X)^i\cdot d$ for $i>3$, which are automatically zero. In other words we have managed to conjugate $D$ into a simpler form that is already decoupled. Thus the cohomology of $D$ and that of $d+Q$ are isomorphic—the isomorphism being given, at the level of fields, by conjugation. So that if $\psi(z)$ is a $d+Q$-cocycle (these are in one-to-one correspondence with $Q$-cocycles as shown above) then $\exp(-\operatorname{ad} X)\cdot \psi(z)$ is a $D$-cocycle. In particular, since conjugation is an automorphism, the operator products are preserved. One still needs to address the picture changing phenomenon. But this is now seen to be simply a feature of the $d$-differential, for which the arguments in the previous subsection apply in full. As a bonus we can find the explicit form of the picture-changing operators $\widetilde{P}_i$ simply by conjugating the original $P_i$.

Although this proof is decidedly shorter and much more elementary; the first proof has its charms too and it moreover applies in situations where the operator $R$ may not be so readily constructed, as evinced by the next section.

## §7 General remarks on the generic embedding

In this section we will consider general features of the embedding between two different BRST cohomology theories. We will prove that for a very general class of embeddings, the analogue of (1.3) is true provided that there is an analogue of (1.2) satisfying some simple properties.

### Embeddings between CFTs

Suppose that we are given two algebras $\mathcal{A}\subset\mathcal{B}$ with basis $\{\phi_i\}$ and $\{\widetilde{\phi}_A\} = \{\widetilde{\phi}_i, \widetilde{\phi}_a\}$, where $\{\widetilde{\phi}_i\}$ generate the $\mathcal{A}$ subalgebra of $\mathcal{B}$. For definiteness we may think of $\mathcal{A}$ and $\mathcal{B}$ as chiral algebras of a conformal field theory, and by generators we mean fields such that all the fields in the chiral algebra are constructed from them via the processes of taking derivatives and/or normal-ordered products. We may assume the existence of an energy-momentum tensor which generates a Virasoro (sub)algebra, but this is not essential. Examples of the algebras we have in mind are the chiral algebras built out of the (super)conformal algebras, affine Lie algebras, $N$=1 affine Lie algebras, and to some extent W-algebras.

Suppose furthermore that we can define BRST operators for both $\mathcal{A}$ and $\mathcal{B}$ and call the respective BRST operators $D_{\mathcal{A}}$ and $D_{\mathcal{B}}$. We will let $\{b_i, c^i\}$ denote the ghosts for $\mathcal{A}$, and $\{b_A, c^A\}$ denote the ghosts for $\mathcal{B}$. Since $\mathcal{A}$ is a subalgebra of $\mathcal{B}$ and we have adapted the generators of $\mathcal{B}$ to this fact, the ghosts are also adapted to this embedding and we have that $\{b_A, c^A\} = \{b_i, b_a, c^i, c^a\}$, where $\{b_i, c^i\}$ are the ghosts for the $\mathcal{A}$ subalgebra and $\{b_a, c^a\}$ are the remaining ghosts.

### The induced module: the ghosts in the machine?

The analogue of the construction (1.2) and (1.3) would be to cook up the fields $\{\widetilde{\phi}_A\}$ starting from the fields $\{\phi_i\}$ and some other "universal" ingredient—that is, something that cannot depend on the precise nature of the $\{\phi_i\}$—in such a way that the BRST cohomology of $D_{\mathcal{A}} = \phi_i c^i + \cdots$ and $D_{\mathcal{B}} = \widetilde{\phi}_A c^A + \cdots$ agree as operator product algebras. The equality of the BRST cohomologies can be understood physically as an equality in the physical spectra of the two theories. This means that the contributions to the physical spectrum of the universal ingredient must cancel precisely the contributions of the extra ghosts $\{b_a, c^a\}$. In particular, as conformal field theories, the universal ingredient must correspond to a CFT with the same degrees of freedom but with opposite (= negative) central charge/level... The reason for this is plain. In the BRST complex for $\mathcal{A}$, which is itself a chiral algebra generated by $\{\phi_i, b_i, c^i\}$, the total generators defined by $\phi_i^{\rm tot} = [D_{\mathcal{A}}, b_i]$ obey $\mathcal{A}$ with zero central element.[1] If we let $\mathsf{C}[\phi]$ denote the contribution of the field(s) $\phi$ to the central element, then we have that the above statement can we written as

$$\sum_i \mathsf{C}[\phi_i] + \sum_i \mathsf{C}[b_i, c^i] = 0 \ . \tag{7.1}$$

Now the same is true for $\mathcal{B}$, whence we have

$$\sum_A \mathsf{C}[\widetilde{\phi}_A] + \sum_A \mathsf{C}[b_A, c^A] = 0 \ . \tag{7.2}$$

Splitting the above sum further we have

$$\sum_i \mathsf{C}[\phi_i] + \mathsf{C}[\text{extra}] + \sum_i \mathsf{C}[b_i, c^i] + \sum_a \mathsf{C}[b_a, c^a] = 0 \ , \tag{7.3}$$

which becomes

$$\mathsf{C}[\text{extra}] = -\sum_a \mathsf{C}[b_a, c^a] = 0 \tag{7.4}$$

after using (7.1).

There is thus a natural candidate for the extra ingredients; namely, BC systems $\{\beta_a, \gamma^a\}$ of the same weights as the extra ghosts but with opposite statistics. For example, in the case of the $N{=}0 \subset N{=}1$ strings, where $\mathcal{A}$ is the Virasoro algebra and $\mathcal{B}$ is the $N{=}1$ superVirasoro algebra, one must add a fermionic BC system of weights $(\frac{3}{2}, -\frac{1}{2})$ to precisely cancel the superconformal ghosts $(\beta, \gamma)$ which are bosonic of the same weights. In the example treated in the preceding sections, we added fermionic BC systems $\{\widetilde{b}_i, \widetilde{c}^i\}$ of weights $(\frac{1}{2}, \frac{1}{2})$ in order to cancel the equally weighted ghosts $\{\beta_i, \gamma^i\}$. In both of these constructions we can think of the extra fields as the ghosts for $\mathcal{A}$ except that their dimensions have been shifted down by $\frac{1}{2}$. This feature—although it seems to have played a guiding rôle in the construction of the $N{=}0 \subset N{=}1$ embedding—has to be considered an artifact of the sort of embedding we are dealing with: namely we are embedding an algebra into its supersymmetrization. In more general embeddings, the fields we must add are sometimes of the same statistics as the generators. One can therefore not confuse them with the ghosts and their identity crisis will be resolved.

[1] Strictly speaking this is only true for Lie (super)algebras. It is well-known that for W-algebras this is not the case simply because the tensor product of two representations—should one exist!—will not be obtained by adding the generators. But nevertheless, in the known examples of BRST complexes for W-algebras, the central element lives already in a Lie subalgebra, and thus we can restrict our attention to those generators.

In both of these examples, the generators $\{\widetilde{\phi}_A\}$ of $\mathcal{B}$ are given in terms of the generators $\{\phi_i\}$ of $\mathcal{A}$ and the extra BC systems $\{\beta_a, \gamma^a\}$ by expressions of the form

$$\begin{aligned} \widetilde{\phi}_i &= \phi_i + \phi_i^{[\beta\gamma]} + \cdots \\ \widetilde{\phi}_a &= \beta_a + \cdots \ , \end{aligned} \tag{7.5}$$

where $\phi_i^{[\beta\gamma]}$ stands for the representation of $\mathcal{A}$ in terms of the $\{\beta_a, \gamma^a\}$ and the $\cdots$ represent corrections by terms of higher $(\beta, \gamma)$-number. Let us call these embeddings "good." In the example treated in the preceding sections, the correction terms were absent, but in the $N{=}0 \subset N{=}1$ embedding the corrections are unavoidable. In this sense, the embedding $\widehat{\mathfrak{g}} \subset S\widehat{\mathfrak{g}}$ is the cleanest of all such constructions. This resulted in the BRST operator $D$ having only $d$ and $\delta$ terms and hence in a double complex. As we will see below, in general we will not have a double complex, but rather a filtered complex and the proof will be slightly more complicated. Happily, though, the isomorphism in cohomology will still be true. In the approach of **[4]**, the simplicity of this example can be seen in the form of the operator $R$ used in the conjugation. In the more general case, and assuming that it exists, $R$ will have a more complicated form which can again be understood as a deformation of the simple $R$ found in section 6 by terms of higher $(\beta, \gamma)$-number.

One should hasten to add that not all embeddings that are known are good. In fact, the only other known example is the case of $N{=}1 \subset N{=}2$ treated in **[1]**, which is not of the form (7.5). But then again it does not seem that the analogue of (1.3) is true in this case. This is not to be thought as an intrinsic property of $N{=}1 \subset N{=}2$ embeddings, but simply of the explicit embedding used in **[1]**. More generally, the existence of the the $\phi_i^{[\beta\gamma]}$ is tied to the fact that the embedding $\mathcal{A} \subset \mathcal{B}$ is "reductive;" that is, that $\mathcal{A}$ has a complement in $\mathcal{B}$ which is a representation of $\mathcal{A}$. This makes sense for $\mathcal{A}$ and $\mathcal{B}$ Lie algebras and agrees with the standard notion of a reductive subalgebra. For W-algebras, we say that $\mathcal{A} \subset \mathcal{B}$ is reductive if in the singular part of the OPEs $\widetilde{\phi}_i(z)\widetilde{\phi}_a(w)$ we only find $\widetilde{\phi}_a$'s. For example, according to this definition, the embedding $\mathsf{Vir} \subset \mathsf{W}_3$ is reductive.

In principle one can try and construct good embeddings of the form (7.5) for any (reductive) $\mathcal{A} \subset \mathcal{B}$ inductively in $(\beta, \gamma)$-number but when one has weight-zero bosons among the extra BC systems, there is no reason a priori that the corrections (either to the $\widetilde{\phi}_A$ or to $R$) should only have a finite number of terms. In fact preliminary investigations with other embeddings suggest that in the generic situation the corrections will involve an infinite number of terms of increasing $(\beta, \gamma)$-number. I have studied the cases of affine Lie algebras $\widehat{\mathfrak{h}} \subset \widehat{\mathfrak{g}}$, $N{=}1 \subset N{=}2$ superconformal algebras, and $\mathsf{Vir} \subset \mathsf{W}_3$ and have managed to obtain the first few terms in the $\cdots$ of (7.5) but have not reached complete expressions. Nevertheless, it seems plausible that without undue effort one should be able to prove inductively that such expressions—however formal—do exist, since modulo the weight-zero bosons there are a finite number of different obstructions. Work towards a cleaner construction of these embeddings is in progress.

A sketch of a proof of (1.3) for good embeddings

Given $\mathcal{A}$ and $\mathcal{B}$, let us nevertheless assume that a good embedding of the form (7.5) does exist. It is then possible to prove that (1.3) is still true; although the proof is perhaps more technical than in the simple example $\widehat{\mathfrak{g}} \subset S\widehat{\mathfrak{g}}$ we considered before. We will furthermore restrict ourselves to the case where $\mathcal{A}$ is a Lie algebra. All interesting examples of which I am aware are of this form and the added complications that could arise in other cases are best handled case by case.

The idea of the proof—as it should be obvious by now—is to assign gradings to our complex in order to break the BRST operator $D_{\mathcal{B}}$ appropriately. Let us assume a good embedding of the form

$$\widetilde{\phi}_i = \sum_{n\geq 0} \widetilde{\phi}_i^{[n]} \qquad \text{and} \qquad \widetilde{\phi}_a = \sum_{n\geq -1} \widetilde{\phi}_a^{[n]} \tag{7.6}$$

where $\widetilde{\phi}_i^{[0]} = \widetilde{\phi}_i + \widetilde{\phi}_i^{[\beta\gamma]}$, $\widetilde{\phi}_i^{[-1]} = \beta_a$ and the grading $[n]$ refers to the $(\beta,\gamma)$-number. It will also prove convenient to keep track of the ghost numbers of the $\{b_i, c^i\}$ and $\{b_a, c^a\}$ separately. To this end we consider the three currents: $J_1 = -b_i c^i$, $J_2 = -b_a c^a$, and $J_3 = -\beta_a\gamma^a$, and we denote their respective charges by $Q_i$. Clearly $Q_1$ and $Q_2$ compute ghost numbers, whereas $Q_3$ takes care of the $(\beta,\gamma)$-number. Let $t \gg 1$ be a large positive integer and consider the combined charge $Q_t = Q_1 + t(Q_2+Q_3)$. Let us decompose the BRST operator $D_{\mathcal{B}}$ into pieces which are homogeneous relative to the grading afforded by $Q_t$. The BRST operator $D_{\mathcal{B}}$ is of the general form:

$$D_{\mathcal{B}} = \sum_{n\geq 0} \widetilde{\phi}_i^{[n]} c^i + \sum_{n\geq -1} \widetilde{\phi}_a^{[n]} c^a - \tfrac{1}{2} f_{AB}{}^C b_C c^A c^B + \cdots$$

where the $f_{AB}{}^C$ could in principle be field dependent (except for the $f_{ij}{}^k$ since we assume that $\mathcal{A}$ is a Lie algebra) and the $\cdots$ stand for terms with quintic and higher order in ghosts, among whom because of reductivity we don't find any $\mathcal{A}$-ghosts.

We now make an extra assumption about the BRST charge $D_{\mathcal{B}}$ (hence independent on the particular embedding) that guarantees that $D_{\mathcal{B}}$ will have only terms of non-negative $Q_t$-charge and that the term of zero charge is linear in the ghosts. For this all we need is that the $f_{ab}{}^c$ and the analogous terms in the $\cdots$ do not depend on the $\{\widetilde{\phi}_a\}$. This is certainly true when $\mathcal{B}$ is a Lie algebra and it is also true for $\mathsf{Vir} \subset \mathsf{W}_3$. It is clear that if this is satisfied, then all terms in $D_{\mathcal{B}}$ have non-negative $Q_t$-charge. Indeed, $\widetilde{\phi}_i^{[n]} c^i$ has charge $1+nt$, $\widetilde{\phi}_a^{[n]} c^a$ has charge $(n+1)t$, $f_{AB}{}^C$ have charge at least 1, whereas the $\cdots$ have charge at least $t$.

Let us write $D_{\mathcal{B}} = \sum_{n\geq 0} d_n$, where $d_n$ has $Q_t$-charge equal to $n$; and let us analyze the $d_n$'s. For $n=0$ we only have $d_0 = \widetilde{\phi}_a^{[-1]} c^a = \beta_a c^a$. For $n=1$ we have $d_1 = \widetilde{\phi}_i^{[0]} c^i - \frac{1}{2} f_{ij}{}^k b_k c^i c^j + f_{ia}{}^b b_b c^a c^i$ which is the BRST operator for $D_{\mathcal{A}}$ but with $\phi_i$ substituted by $\phi_i + \phi_i^{[\beta\gamma]} + \phi_i^{[bc]}$, where $\phi_i^{[bc]}$ are the generators of $\mathcal{A}$ in the representation afforded by the $\{b_a, c^a\}$ ghosts. It is moreover clear that the next $d_n$ will be precisely $d_t$. Since the $Q_t$ charge is additive over the operator product algebra, the equation $D_{\mathcal{B}}^2 = 0$ breaks up into homogeneous pieces. The first few three equations are $d_0^2 = d_1^2 = [d_0, d_1] = 0$. By refining the $Q_t$-charge further, we can define a bigrading, relative to which $d_0$ has degree $(0,1)$ and $d_1$ has degree $(1,0)$ whence, as before, we have a double complex. A similar analysis to that of the preceding sections shows that $d_0$ only has cohomology when the occupation numbers of $\{b_a, c^a\}$ and $\{\beta_a, \gamma_a\}$ are separately zero. One can go down the descent equations again and recover the result that the cohomology of $d_0 + d_1$ is isomorphic to the cohomology of the operator $D_{\mathcal{A}}$. But how about the other terms in $D_{\mathcal{B}}$? One can see—at least in the examples of interest—that if we write $D_{\mathcal{B}} = d + \delta$, with $d = d_0 + d_1$, then again $d^2 = \delta^2 = [d, \delta] = 0$ and there exists a choice of bigrading refining the $Q_t$-charge in such a way that we have again a double complex. Using again that the cohomology of $d$ is trivial unless the occupation numbers of $\{b_a, c^a\}$ and $\{\beta_a, \gamma_a\}$ are separately zero, and noticing that all terms in $\delta$ have nonzero occupation numbers in these variables, we conclude that the $D_{\mathcal{B}}$-cocycles are in one-to-one correspondence with $d$-cocycles, whence with $D_{\mathcal{A}}$-cocycles. The same correspondences exist between the coboundaries, and as a result the cohomologies are isomorphic. Moreover the isomorphism extends to one of operator product algebras, as the proof of this fact (see section 5) only depended on manipulating the descent equations, which we have here as well: in fact, we have two sets of descent equations since we have written the BRST complex of $D_{\mathcal{B}}$ as two iterated double complexes, each with its descent equations. The discussion of the picture changing operators again goes through *mutatis mutandis*.

One again could wonder whether this could be proven in a much simpler way via a suitable $R$ operator by which to "dress" the simpler differential $d_0 + D_{\mathcal{A}}$ to obtain $D_{\mathcal{B}}$. Let us briefly comment on this. Our results above show that under the hypothesis stated, the cohomologies of $D_{\mathcal{B}}$ and $d_0 + D_{\mathcal{A}}$ are isomorphic. Hence we are in the case that two differentials acting on the same space have identical cohomolgy. Can one conclude that there is a map relating the two differentials? In this case, since the complexes are actually vector spaces, a *linear* map intertwining between the two differentials can always be constructed. Hence we can lift any map in cohomology to a chain map between complexes. However, this is not enough. We are not simply after a linear map, but rather after an operator algebra homomorphism. It would be interesting to understand the conditions on our algebras such that this is possible. It may be that under sufficiently general assumptions, like the ones that allowed us to prove the more general version of (1.3), such a map can be found rendering the isomorphism of the cohomologies as operator product algebras manifest.

## ACKNOWLEDGEMENTS

Some of the ideas in the present paper crystallized as a result of electronic communications with Takashi Kimura, Jim Stasheff, and Alexander Voronov. It is a pleasure to thank them for their stimulating correspondence; as it is also to thank Nathan Berkovits, Michael Green, Chris Hull, Eduardo Ramos, and Sonia Stanciu for helpful conversations.